\documentclass[sigplan,screen,sigconf]{acmart}
\usepackage{soul}

\AtBeginDocument{%
  \providecommand\BibTeX{{%
    \normalfont B\kern-0.5em{\scshape i\kern-0.25em b}\kern-0.8em\TeX}}}

\setcopyright{rightsretained}
\acmDOI{}

\acmISBN{}
\acmConference[LATTE '21]{1st Workshop on Languages, Tools, and Techniques for Accelerator Design}{April 15, 2021}{Virtual, Earth}

\usepackage{xspace}
\usepackage{caption}
\usepackage{subcaption}
\usepackage[frozencache=true,cachedir=minted-cache]{minted}
\usepackage{listings}
\usepackage{color, colortbl}
\usepackage{tikz}

\lstdefinelanguage{llvm}{
    alsoletter={\%,\#,!},
    keywordsprefix={\%,\#},
    keywordstyle=\color{orange3},
    otherkeywords={index, f32, i32, f64, i8},
    keywords=[3]{index, f32, memref, i32, f64, i8, max, min, affine_map, affine_set, to, for, parallel_for, load, store, addf},
    keywordstyle=[3]\color{scarletred3},
    showstringspaces=false,
    breaklines=true,
    breakatwhitespace=true,
    morestring=[b]",
    stringstyle=\color{green3},
    moredelim=[s][\color{scarletred3}]{!}{\ },
}

\definecolor{orange3}{rgb}{0.808,0.361,0.000}
\definecolor{scarletred3}{rgb}{0.643,0.000,0.000}
\definecolor{green3}{rgb}{0.000,0.405,0.000}
\definecolor{blue3}{rgb}{0.000,0.000,0.704}
\definecolor{aluminium1}{rgb}{0.933,0.933,0.925}
\definecolor{aluminium2}{rgb}{0.827,0.843,0.812}
\definecolor{aluminium3}{rgb}{0.729,0.741,0.714}
\definecolor{aluminium4}{rgb}{0.533,0.541,0.522}
\definecolor{aluminium5}{rgb}{0.333,0.341,0.325}
\definecolor{aluminium6}{rgb}{0.180,0.204,0.212}

\newcommand{\tool}{$\phi_{sm}$\xspace}

\begin{document}

\title{Phism: Polyhedral High-Level Synthesis in MLIR}
\subtitle{(Work in Progress)}


\author{Ruizhe Zhao}
\email{ruizhe.zhao15@imperial.ac.uk}
\affiliation{
  \institution{Imperial College London}
  \country{UK}
}
\author{Jianyi Cheng}
\email{jianyi.cheng17@imperial.ac.uk}
\affiliation{
  \institution{Imperial College London}
  \country{UK}
}


\begin{abstract}
  Polyhedral optimisation, a methodology that views nested loops as polyhedra and searches for their optimal transformation regarding specific objectives (parallelism, locality, etc.), sounds promising for mitigating difficulties in automatically optimising hardware designs described by high-level synthesis (HLS), which are typically software programs with nested loops. Nevertheless, existing polyhedral tools cannot meet the requirements from HLS developers for platform-specific customisation and software/hardware co-optimisation. This paper proposes \tool (Phism), a polyhedral HLS framework built on MLIR, to address these challenges through progressive lowering multi-level intermediate representations (IRs) from polyhedra to HLS designs.
\end{abstract}



\keywords{}

\maketitle

\section{Introduction}\label{sec:intro}

\begin{figure}[t]
  \centering
  \includegraphics[width=0.75\columnwidth]{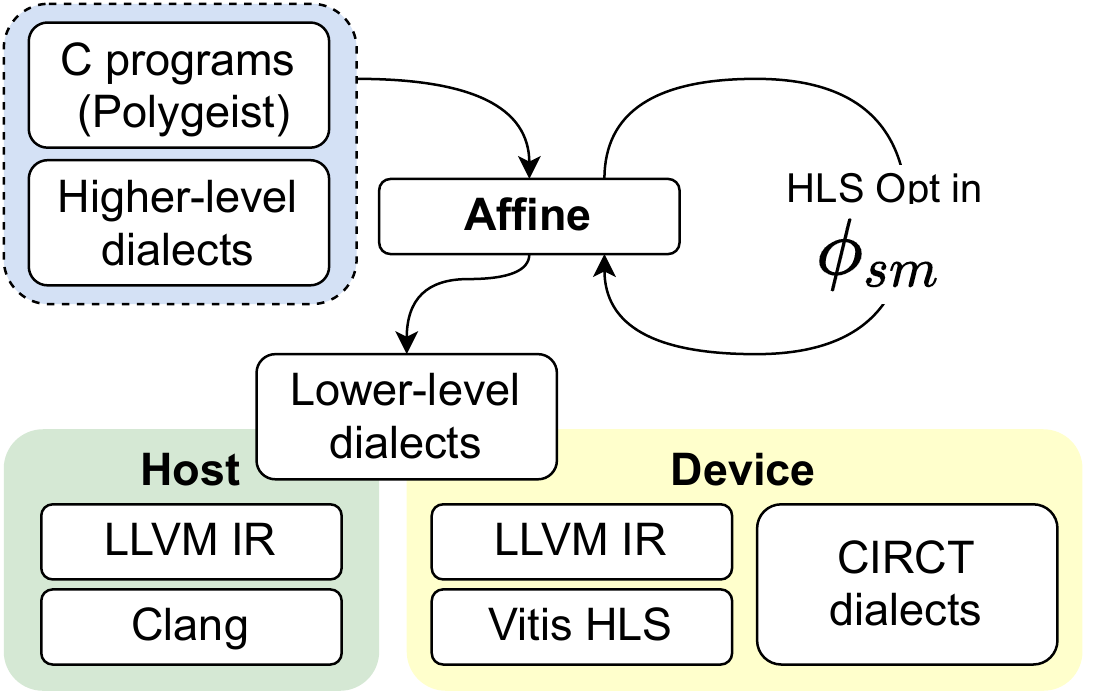}
  \caption{\tool overview. Polygeist~\cite{polygeist} provides both the conversion from C and the polyhedral optimisation on Affine.}
  \label{fig:overview}
\end{figure}

High-level synthesis~(HLS) can transform software programs in C-like languages into hardware designs, and polyhedral optimisation can provide elegant solutions for various problems in this process, e.g., loop pipelining and splitting~\cite{liu2016loop,liu2017polyhedral}, loop transformations~\cite{zuo2013improving}, design generation~\cite{cong2018polysa,wang2021autosa}, etc. It is mainly due to many HLS programs, originally described in C-like languages, have regions with control flow and dependence relations that can be formulated as affine expressions at compile time. These regions, conventionally referred to as Static Control Parts (SCoPs) by polyhedral research, are where polyhedral optimisation can be fully leveraged.

Nevertheless, existing polyhedral tools are incapable of keeping up with the recent progress in HLS: more target platforms, design models~\cite{CIRCT,josipovic2018dynamically}, and applications are being supported, while polyhedral tools, e.g., isl~\cite{verdoolaege2010isl} and Pluto~\cite{bondhugula2008practical}, are not versatile enough to be customised for them. Existing papers~\cite{zuo2013codegen,PoTHoLeS} can partially address this challenge by customising some processing stages when lowering from polyhedral representations, but we need a more comprehensive approach to meet the current and future demands.

Inspired by recent work on domain-specific compiler~\cite{gysi2020domain} and hardware synthesis~\cite{CIRCT}, we are motivated to {\em progressively lower} from polyhedra to HLS designs, so that we can customise each processing stage at its right abstraction level and define composable and reusable transformations for future extension. MLIR~\cite{lattner2020mlir} is a perfect fit for our objectives as a compiler infrastructure that effectively supports multi-level intermediate representations (IRs) definition and transformation, under the concept of {\em dialect}. Therefore, this paper proposes \tool (Figure~\ref{fig:overview}), the first MLIR-based polyhedral HLS tool featuring progressively lowering by:

\begin{enumerate}
  \item implementing HLS optimisation at the right abstraction levels during progressively lowering;
  \item leveraging dialects, e.g., Affine~\cite{MLIRAffine}, to build transformations for HLS on polyhedral representations;
  \item connecting with various sources, e.g., C or other higher-level dialects, and targets, e.g., Vitis~\cite{kathail2020xilinx} or CIRCT~\cite{CIRCT}.
\end{enumerate}

In this way, \tool can better leverage polyhedral optimisation for HLS, and therefore, provide an efficient hardware design method.

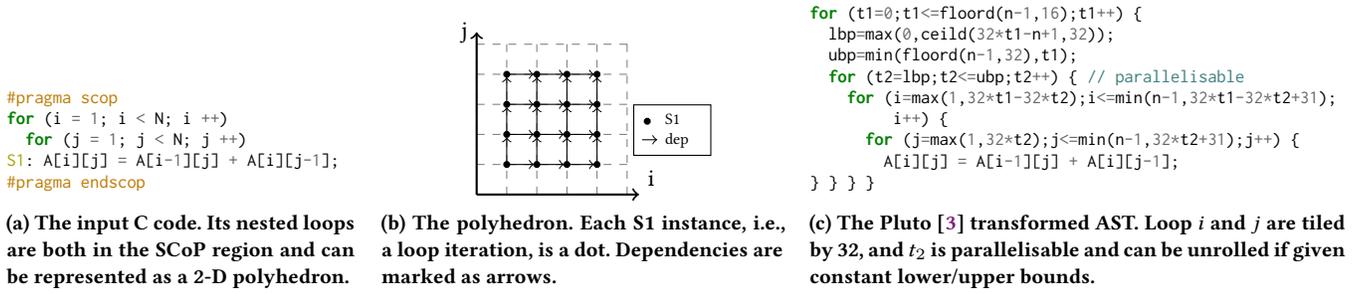
\begin{figure*}[t]
  \centering
  \begin{subfigure}[b]{0.26\textwidth}
    \centering
    \begin{minted}[fontsize=\footnotesize]{c}
#pragma scop
for (i = 1; i < N; i ++)
  for (j = 1; j < N; j ++)
S1: A[i][j] = A[i-1][j] + A[i][j-1];
#pragma endscop
    \end{minted}
    \caption{The input C code. Its nested loops are both in the SCoP region and can be represented as a 2-D polyhedron.}
    \label{fig:inputprog}
  \end{subfigure}\hfill
  \begin{subfigure}[b]{0.30\textwidth}
    \centering
    \begin{tikzpicture}[
        scale=0.8,
        roundnode/.style={circle, draw=black, fill=black, very thick, minimum size=0.5mm, inner sep=0pt},
        legendtext/.style={text width=5mm, inner sep=0pt}
      ]
      \draw[color=gray,step=.5cm, dashed] (0,0) grid (2.5,2.5);
      \draw[->,thick] (0,0) -- (2.7,0) node[above right] {i};
      \draw[->,thick] (0,0) -- (0,2.7) node[left] {j};
      \foreach \i in {0.5,1,...,2}
      \foreach \j in {0.5,1,...,2}
      \node[roundnode] at (\i,\j) {};
      \foreach \i in {0.5,1,...,1.5}
      \foreach \j in {0.5,1,...,2} {
          \draw[->] (\i+0.05,\j) -- (\i+0.45,\j);
          \draw[->] (\j,\i+0.05) -- (\j,\i+0.45);
        }
      \matrix [draw,below left,nodes={font=\scriptsize},column sep=1mm,row sep=1mm] at (3.9,1.5) {
        \node [roundnode] at (0.1,-0.08) {}; & \node [legendtext] {S1};  \\
        \draw[->] (0,-0.1) -- (0.2,-0.1); & \node[legendtext] {dep} ; \\
      };
    \end{tikzpicture}
    \caption{The polyhedron. Each S1 instance, i.e., a loop iteration, is a dot. Dependencies are marked as arrows.}
    \label{fig:origpoly}
  \end{subfigure}\hfill
  \begin{subfigure}[b]{0.40\textwidth}
    \centering
    \begin{minted}[fontsize=\footnotesize]{c}
for (t1=0;t1<=floord(n-1,16);t1++) {
  lbp=max(0,ceild(32*t1-n+1,32));
  ubp=min(floord(n-1,32),t1);
  for (t2=lbp;t2<=ubp;t2++) { // parallelisable
    for (i=max(1,32*t1-32*t2);i<=min(n-1,32*t1-32*t2+31);
         i++) {
      for (j=max(1,32*t2);j<=min(n-1,32*t2+31);j++) {
        A[i][j] = A[i-1][j] + A[i][j-1];
} } } }
    \end{minted}
    \caption{The Pluto~\cite{bondhugula2008practical} transformed AST. Loop $i$ and $j$ are tiled by 32, and $t_2$ is parallelisable and can be unrolled if given constant lower/upper bounds.}
    \label{fig:transprog}
  \end{subfigure}

  \caption{Some representations of an example program~\cite{bondhugula2008effective} that can be polyhedral transformed by Pluto~\cite{bondhugula2008practical}.}
  \label{fig:example}
\end{figure*}

\section{Background and Related Work}

\paragraph{Polyhedral optimisation}
There are decades of research on representing programs in polyhedra~\cite{feautrier1992some} and transforming them for better performance~\cite{bondhugula2008practical}. Polyhedral optimisation transforms polyhedra extracted from original programs, and the resulting polyhedra should be converted back through non-trivial polyhedral code generation~\cite{bastoul2004code,chen2012polyhedra,grosser2015polyhedral} for various platforms~\cite{zuo2013codegen,verdoolaege2017scheduling}.

\paragraph{MLIR} MLIR is a compiler infrastructure for building IRs and their transformations~\cite{lattner2020mlir}. Here, IRs are in static single assignment (SSA)~\cite{rosen1988global} forms, and MLIR provides the {\em dialect} mechanism to define IRs for domain-specific problems within the ecosystem~\cite{MLIRLangRef}. This paper focuses on Affine~\cite{MLIRAffine}, which is designed for representing polyhedral programs, and can be translated from C and optimised by existing polyhedral tools enabled by Polygeist~\cite{polygeist}.

\paragraph{Related work} PoTHoLeS~\cite{PoTHoLeS} provides a polyhedral compilation tool for HLS, and Zuo et al.~\cite{zuo2013codegen} describe several polyhedral code generation techniques for HLS. \tool is compatible with these prior approaches and is more versatile to cover recent HLS techniques and software/hardware co-optimisation with progressive lowering.

\section{Overview}\label{sec:method}

\tool aims to carry out polyhedral HLS at the right abstraction level for progressive lowering stages. Here, we first demonstrate the abstraction levels that existing tools can provide for a concrete example, and why they are insufficient for the growing demands from polyhedral HLS applications. Next, we discuss what new abstraction layers that \tool can introduce to improve polyhedral HLS.

\subsection{Limitations in Existing Tools}

Using existing tools like Pluto~\cite{bondhugula2008practical} and CLooG~\cite{bastoul2004code}, we can represent the input C code in Figure~\ref{fig:inputprog}, which has an annotated SCoP region, as a polyhedron (Figure~\ref{fig:origpoly}) and then optimise it into a C-like AST form as in Figure~\ref{fig:transprog}.
There is a big gap between this product and what an HLS tool normally expect, specifically:

\begin{enumerate}
  \item It is uncertain which loops should be placed on hardware.
  \item Optimisation directives, e.g., pipelining, unrolling, etc. should be inserted to achieve higher performance.
  \item Details about host interacts with the hardware, e.g., data transfer patterns, are opaque.
\end{enumerate}

Unfortunately, abstraction layers from existing polyhedral tools hinder us from filling the gap. There are only two representations available, {\em polyhedron representation} and {\em C-like AST}.
Polyhedron representations can encode iteration domain, schedule, and memory access as matrices~\cite{girbal2006semi}, but they are too abstract to describe HLS optimisation. An HLS optimisation normally views its input as concrete loops and conditionals, which are unknown from the polyhedron representation unless we export it using an extra, irreversible code generator~\cite{bastoul2004code}, e.g., a single statement in a stencil-based computation can be represented by a single polyhedron, but there can be hundreds of loops and conditionals generated from that to deal with various boundaries.
C-like ASTs, on the other hand, are too close to what HLS tools take, any higher-level optimisation opportunities may already be missed at this low abstraction level since polyhedral information are not there anymore. Only non-polyhedral source-to-source transformation is possible.

\begin{figure}[h]
  \centering
  {\scriptsize
    \begin{lstlisting}[language=llvm]
    #map0 = affine_map<()[s0] -> ((s0-1) floordiv 16 + 1)>
    #map1 = affine_map<(d0)[s0] -> (0, (d0*32-s0+1) ceildiv 32)>
    #map2 = affine_map<(d0)[s0] -> ((s0-1) floordiv 32 + 1, d0+1)>
    #map3 = affine_map<(d0,d1) -> (1, d0*32-d1*32)>
    #map4 = affine_map<(d0,d1)[s0] -> (s0, d0*32-d1*32+32)>
    #map5 = affine_map<(d0) -> (1, d0*32)>
    #map6 = affine_map<(d0)[s0] -> (s0, d0*32+32)>
    affine.for %t1 = 0 to #map0()[%N] {
      affine.parallel_for %t2 = max #map1(%t1)[%N] to
                                min #map2(%t1)[%N] {
        affine.for %i = max #map3(%t1, %t2) to
                        min #map4(%t1, %t2)[%N] {
          affine.for %j = max #map5(%t2) to min #map6(%t2)[%N] {
            call @S1(%A, %i, %j) 
    } } } }
  \end{lstlisting}
  }
  \caption{An Affine representation of the Pluto-transformed code in Figure~\ref{fig:transprog} from~\cite{polygeist}. Syntax details are in~\cite{MLIRAffine}.}
  \label{fig:affine}
\end{figure}

\subsection{Our Approach}

\tool aims to fill the gap through progressive lowering in MLIR, specifically, using the Affine dialect~\cite{MLIRAffine}.
Affine can perfectly address the aforementioned issues in existing tools: Affine describes polyhedron that polyhedral transformation can work on, and it has loops and conditionals for us to describe HLS optimisation.
For example, Figure~\ref{fig:affine} shows the Affine code equivalent to the C-like AST produced by Pluto (Figure~\ref{fig:transprog}), and since Affine restricts that its loops are bounded by affine combinations, this code piece also describes the transformed polyhedron.

The {\em sub-bounding-box tiling} algorithm~\cite{zuo2013codegen}, which unifies the tile bounds for uniform workload distribution among processing units, is a perfect example showing the advantage of \tool leveraging Affine. Its original implementation needs to reproduce polyhedra from CLooG-generated code to find parallelogram hulls and regenerate C code in the end, while using Affine, we can calculate the hulls and transform the code directly in the same representation, which is more efficient, less error-prone, and easier to integrate with precedent and subsequent transformations.

After Affine, we can progressively lower the abstraction level to other dialects. During this procedure, we can describe software/hardware partition, design space exploration, data layout optimisation, and many other techniques as MLIR transformations.
Once we reach Standard~\cite{MLIRStandard}, the dialect at a level right above LLVM IR, we can decide whether export to Vitis~\cite{kathail2020xilinx}, or continue lowering to hardware description dialects in CIRCT~\cite{CIRCT} (Figure~\ref{fig:overview}), to finally produce an accelerator design.

\section{Summary}

This paper presents the concepts of \tool, a polyhedral HLS tool built upon MLIR adapting progressive lowering. \tool can narrow the gap between polyhedral representation and HLS optimisation by lowering from the MLIR Affine dialect and transforming IRs at the right abstraction levels. More details and the current progress can be found in \url{https://github.com/kumasento/polymer}.


\newpage
\bibliographystyle{ACM-Reference-Format}
\bibliography{ref}


\begin{thebibliography}{26}


\ifx \showCODEN    \undefined \def \showCODEN     #1{\unskip}     \fi
\ifx \showDOI      \undefined \def \showDOI       #1{#1}\fi
\ifx \showISBNx    \undefined \def \showISBNx     #1{\unskip}     \fi
\ifx \showISBNxiii \undefined \def \showISBNxiii  #1{\unskip}     \fi
\ifx \showISSN     \undefined \def \showISSN      #1{\unskip}     \fi
\ifx \showLCCN     \undefined \def \showLCCN      #1{\unskip}     \fi
\ifx \shownote     \undefined \def \shownote      #1{#1}          \fi
\ifx \showarticletitle \undefined \def \showarticletitle #1{#1}   \fi
\ifx \showURL      \undefined \def \showURL       {\relax}        \fi
\providecommand\bibfield[2]{#2}
\providecommand\bibinfo[2]{#2}
\providecommand\natexlab[1]{#1}
\providecommand\showeprint[2][]{arXiv:#2}

\bibitem[\protect\citeauthoryear{Bastoul}{Bastoul}{2004}]%
        {bastoul2004code}
\bibfield{author}{\bibinfo{person}{C{\'e}dric Bastoul}.}
  \bibinfo{year}{2004}\natexlab{}.
\newblock \showarticletitle{Code generation in the polyhedral model is easier
  than you think}. In \bibinfo{booktitle}{\emph{{PACT}}}. IEEE,
  \bibinfo{pages}{7--16}.
\newblock


\bibitem[\protect\citeauthoryear{Bayliss}{Bayliss}{2014}]%
        {PoTHoLeS}
\bibfield{author}{\bibinfo{person}{Samuel Bayliss}.}
  \bibinfo{year}{2014}\natexlab{}.
\newblock \bibinfo{title}{{PoTHoLeS: Polyhedral Compilation Tool for High Level
  Synthesis}}.
\newblock
\newblock
\urldef\tempurl%
\url{https://github.com/SamuelBayliss/Potholes}
\showURL{%
\tempurl}


\bibitem[\protect\citeauthoryear{Bondhugula, Hartono, Ramanujam, and
  Sadayappan}{Bondhugula et~al\mbox{.}}{2008}]%
        {bondhugula2008practical}
\bibfield{author}{\bibinfo{person}{Uday Bondhugula}, \bibinfo{person}{Albert
  Hartono}, \bibinfo{person}{Jagannathan Ramanujam}, {and}
  \bibinfo{person}{Ponnuswamy Sadayappan}.} \bibinfo{year}{2008}\natexlab{}.
\newblock \showarticletitle{A practical automatic polyhedral parallelizer and
  locality optimizer}. In \bibinfo{booktitle}{\emph{{PLDI}}}.
  \bibinfo{pages}{101--113}.
\newblock


\bibitem[\protect\citeauthoryear{Bondhugula}{Bondhugula}{2008}]%
        {bondhugula2008effective}
\bibfield{author}{\bibinfo{person}{Uday~Kumar Bondhugula}.}
  \bibinfo{year}{2008}\natexlab{}.
\newblock \emph{\bibinfo{title}{Effective automatic parallelization and
  locality optimization using the polyhedral model}}.
\newblock \bibinfo{thesistype}{Ph.D. Dissertation}. \bibinfo{school}{The Ohio
  State University}.
\newblock


\bibitem[\protect\citeauthoryear{Chen}{Chen}{2012}]%
        {chen2012polyhedra}
\bibfield{author}{\bibinfo{person}{Chun Chen}.}
  \bibinfo{year}{2012}\natexlab{}.
\newblock \showarticletitle{Polyhedra scanning revisited}. In
  \bibinfo{booktitle}{\emph{Proceedings of the 33rd ACM SIGPLAN conference on
  Programming Language Design and Implementation}}. \bibinfo{pages}{499--508}.
\newblock


\bibitem[\protect\citeauthoryear{{CIRCT}}{{CIRCT}}{[n.d.]}]%
        {CIRCT}
\bibfield{author}{\bibinfo{person}{{CIRCT}}.}
  \bibinfo{year}{[n.d.]}\natexlab{}.
\newblock \bibinfo{title}{{``CIRCT'' / Circuit IR Compilers and Tools}}.
\newblock \bibinfo{howpublished}{\url{https://github.com/llvm/circt}}.
\newblock


\bibitem[\protect\citeauthoryear{Cong and Wang}{Cong and Wang}{2018}]%
        {cong2018polysa}
\bibfield{author}{\bibinfo{person}{Jason Cong} {and} \bibinfo{person}{Jie
  Wang}.} \bibinfo{year}{2018}\natexlab{}.
\newblock \showarticletitle{PolySA: Polyhedral-based systolic array
  auto-compilation}. In \bibinfo{booktitle}{\emph{{ICCAD}}}. IEEE,
  \bibinfo{pages}{1--8}.
\newblock


\bibitem[\protect\citeauthoryear{developers}{developers}{[n.d.]}]%
        {MLIRLangRef}
\bibfield{author}{\bibinfo{person}{{MLIR} developers}.}
  \bibinfo{year}{[n.d.]}\natexlab{}.
\newblock \bibinfo{title}{MLIR Language Reference}.
\newblock \bibinfo{howpublished}{\url{https://mlir.llvm.org/docs/LangRef/}}.
\newblock


\bibitem[\protect\citeauthoryear{Feautrier}{Feautrier}{1992}]%
        {feautrier1992some}
\bibfield{author}{\bibinfo{person}{Paul Feautrier}.}
  \bibinfo{year}{1992}\natexlab{}.
\newblock \showarticletitle{Some efficient solutions to the affine scheduling
  problem. Part II. Multidimensional time}.
\newblock \bibinfo{journal}{\emph{International journal of parallel
  programming}} \bibinfo{volume}{21}, \bibinfo{number}{6}
  (\bibinfo{year}{1992}), \bibinfo{pages}{389--420}.
\newblock


\bibitem[\protect\citeauthoryear{Girbal, Vasilache, Bastoul, Cohen, Parello,
  Sigler, and Temam}{Girbal et~al\mbox{.}}{2006}]%
        {girbal2006semi}
\bibfield{author}{\bibinfo{person}{Sylvain Girbal}, \bibinfo{person}{Nicolas
  Vasilache}, \bibinfo{person}{C{\'e}dric Bastoul}, \bibinfo{person}{Albert
  Cohen}, \bibinfo{person}{David Parello}, \bibinfo{person}{Marc Sigler}, {and}
  \bibinfo{person}{Olivier Temam}.} \bibinfo{year}{2006}\natexlab{}.
\newblock \showarticletitle{Semi-automatic composition of loop transformations
  for deep parallelism and memory hierarchies}.
\newblock \bibinfo{journal}{\emph{International Journal of Parallel
  Programming}} \bibinfo{volume}{34}, \bibinfo{number}{3}
  (\bibinfo{year}{2006}), \bibinfo{pages}{261--317}.
\newblock


\bibitem[\protect\citeauthoryear{Grosser, Verdoolaege, and Cohen}{Grosser
  et~al\mbox{.}}{2015}]%
        {grosser2015polyhedral}
\bibfield{author}{\bibinfo{person}{Tobias Grosser}, \bibinfo{person}{Sven
  Verdoolaege}, {and} \bibinfo{person}{Albert Cohen}.}
  \bibinfo{year}{2015}\natexlab{}.
\newblock \showarticletitle{Polyhedral AST generation is more than scanning
  polyhedra}.
\newblock \bibinfo{journal}{\emph{ACM Transactions on Programming Languages and
  Systems (TOPLAS)}} \bibinfo{volume}{37}, \bibinfo{number}{4}
  (\bibinfo{year}{2015}), \bibinfo{pages}{1--50}.
\newblock


\bibitem[\protect\citeauthoryear{Gysi, M{\"u}ller, Zinenko, Herhut, Davis,
  Wicky, Fuhrer, Hoefler, and Grosser}{Gysi et~al\mbox{.}}{2020}]%
        {gysi2020domain}
\bibfield{author}{\bibinfo{person}{Tobias Gysi}, \bibinfo{person}{Christoph
  M{\"u}ller}, \bibinfo{person}{Oleksandr Zinenko}, \bibinfo{person}{Stephan
  Herhut}, \bibinfo{person}{Eddie Davis}, \bibinfo{person}{Tobias Wicky},
  \bibinfo{person}{Oliver Fuhrer}, \bibinfo{person}{Torsten Hoefler}, {and}
  \bibinfo{person}{Tobias Grosser}.} \bibinfo{year}{2020}\natexlab{}.
\newblock \showarticletitle{Domain-specific Multi-Level IR rewriting for GPU}.
\newblock \bibinfo{journal}{\emph{arXiv preprint arXiv:2005.13014}}
  (\bibinfo{year}{2020}).
\newblock


\bibitem[\protect\citeauthoryear{Josipovi{\'c}, Ghosal, and
  Ienne}{Josipovi{\'c} et~al\mbox{.}}{2018}]%
        {josipovic2018dynamically}
\bibfield{author}{\bibinfo{person}{Lana Josipovi{\'c}},
  \bibinfo{person}{Radhika Ghosal}, {and} \bibinfo{person}{Paolo Ienne}.}
  \bibinfo{year}{2018}\natexlab{}.
\newblock \showarticletitle{Dynamically scheduled high-level synthesis}. In
  \bibinfo{booktitle}{\emph{Proceedings of the 2018 ACM/SIGDA International
  Symposium on Field-Programmable Gate Arrays}}. \bibinfo{pages}{127--136}.
\newblock


\bibitem[\protect\citeauthoryear{Kathail}{Kathail}{2020}]%
        {kathail2020xilinx}
\bibfield{author}{\bibinfo{person}{Vinod Kathail}.}
  \bibinfo{year}{2020}\natexlab{}.
\newblock \showarticletitle{Xilinx Vitis unified software platform}. In
  \bibinfo{booktitle}{\emph{Proceedings of the 2020 ACM/SIGDA International
  Symposium on Field-Programmable Gate Arrays}}. \bibinfo{pages}{173--174}.
\newblock


\bibitem[\protect\citeauthoryear{Lattner, Pienaar, Amini, Bondhugula, Riddle,
  Cohen, Shpeisman, Davis, Vasilache, and Zinenko}{Lattner
  et~al\mbox{.}}{2020}]%
        {lattner2020mlir}
\bibfield{author}{\bibinfo{person}{Chris Lattner}, \bibinfo{person}{Jacques
  Pienaar}, \bibinfo{person}{Mehdi Amini}, \bibinfo{person}{Uday Bondhugula},
  \bibinfo{person}{River Riddle}, \bibinfo{person}{Albert Cohen},
  \bibinfo{person}{Tatiana Shpeisman}, \bibinfo{person}{Andy Davis},
  \bibinfo{person}{Nicolas Vasilache}, {and} \bibinfo{person}{Oleksandr
  Zinenko}.} \bibinfo{year}{2020}\natexlab{}.
\newblock \showarticletitle{MLIR: A compiler infrastructure for the end of
  Moore's law}.
\newblock \bibinfo{journal}{\emph{arXiv preprint arXiv:2002.11054}}
  (\bibinfo{year}{2020}).
\newblock


\bibitem[\protect\citeauthoryear{Liu, Wickerson, Bayliss, and
  Constantinides}{Liu et~al\mbox{.}}{2017}]%
        {liu2017polyhedral}
\bibfield{author}{\bibinfo{person}{Junyi Liu}, \bibinfo{person}{John
  Wickerson}, \bibinfo{person}{Samuel Bayliss}, {and} \bibinfo{person}{George~A
  Constantinides}.} \bibinfo{year}{2017}\natexlab{}.
\newblock \showarticletitle{Polyhedral-based dynamic loop pipelining for
  high-level synthesis}.
\newblock \bibinfo{journal}{\emph{IEEE Transactions on Computer-Aided Design of
  Integrated Circuits and Systems}} \bibinfo{volume}{37}, \bibinfo{number}{9}
  (\bibinfo{year}{2017}), \bibinfo{pages}{1802--1815}.
\newblock


\bibitem[\protect\citeauthoryear{Liu, Wickerson, and Constantinides}{Liu
  et~al\mbox{.}}{2016}]%
        {liu2016loop}
\bibfield{author}{\bibinfo{person}{Junyi Liu}, \bibinfo{person}{John
  Wickerson}, {and} \bibinfo{person}{George~A Constantinides}.}
  \bibinfo{year}{2016}\natexlab{}.
\newblock \showarticletitle{Loop splitting for efficient pipelining in
  high-level synthesis}. In \bibinfo{booktitle}{\emph{{FCCM}}}. IEEE,
  \bibinfo{pages}{72--79}.
\newblock


\bibitem[\protect\citeauthoryear{{MLIR}}{{MLIR}}{[n.d.]a}]%
        {MLIRAffine}
\bibfield{author}{\bibinfo{person}{{MLIR}}.}
  \bibinfo{year}{[n.d.]}\natexlab{a}.
\newblock \bibinfo{title}{{`affine' Dialect}}.
\newblock
  \bibinfo{howpublished}{\url{https://mlir.llvm.org/docs/Dialects/Affine/}}.
\newblock


\bibitem[\protect\citeauthoryear{{MLIR}}{{MLIR}}{[n.d.]b}]%
        {MLIRStandard}
\bibfield{author}{\bibinfo{person}{{MLIR}}.}
  \bibinfo{year}{[n.d.]}\natexlab{b}.
\newblock \bibinfo{title}{{`std' Dialect}}.
\newblock
  \bibinfo{howpublished}{\url{https://mlir.llvm.org/docs/Dialects/Standard/}}.
\newblock


\bibitem[\protect\citeauthoryear{Moses, Chelini, Zhao, and Zinenko}{Moses
  et~al\mbox{.}}{2021}]%
        {polygeist}
\bibfield{author}{\bibinfo{person}{William~S Moses}, \bibinfo{person}{Lorenzo
  Chelini}, \bibinfo{person}{Ruizhe Zhao}, {and} \bibinfo{person}{Oleksandr
  Zinenko}.} \bibinfo{year}{2021}\natexlab{}.
\newblock \showarticletitle{Polygeist: Affine C in MLIR}. In
  \bibinfo{booktitle}{\emph{IMPACT}}.
\newblock


\bibitem[\protect\citeauthoryear{Rosen, Wegman, and Zadeck}{Rosen
  et~al\mbox{.}}{1988}]%
        {rosen1988global}
\bibfield{author}{\bibinfo{person}{Barry~K Rosen}, \bibinfo{person}{Mark~N
  Wegman}, {and} \bibinfo{person}{F~Kenneth Zadeck}.}
  \bibinfo{year}{1988}\natexlab{}.
\newblock \showarticletitle{Global value numbers and redundant computations}.
  In \bibinfo{booktitle}{\emph{Proceedings of the 15th ACM SIGPLAN-SIGACT
  symposium on Principles of programming languages}}. \bibinfo{pages}{12--27}.
\newblock


\bibitem[\protect\citeauthoryear{Verdoolaege}{Verdoolaege}{2010}]%
        {verdoolaege2010isl}
\bibfield{author}{\bibinfo{person}{Sven Verdoolaege}.}
  \bibinfo{year}{2010}\natexlab{}.
\newblock \showarticletitle{isl: An integer set library for the polyhedral
  model}. In \bibinfo{booktitle}{\emph{International Congress on Mathematical
  Software}}. Springer, \bibinfo{pages}{299--302}.
\newblock


\bibitem[\protect\citeauthoryear{Verdoolaege and Janssens}{Verdoolaege and
  Janssens}{2017}]%
        {verdoolaege2017scheduling}
\bibfield{author}{\bibinfo{person}{Sven Verdoolaege} {and}
  \bibinfo{person}{Gerda Janssens}.} \bibinfo{year}{2017}\natexlab{}.
\newblock \showarticletitle{Scheduling for PPCG}.
\newblock \bibinfo{journal}{\emph{Report CW}}  \bibinfo{volume}{706}
  (\bibinfo{year}{2017}).
\newblock


\bibitem[\protect\citeauthoryear{Wang, Guo, and Cong}{Wang
  et~al\mbox{.}}{2021}]%
        {wang2021autosa}
\bibfield{author}{\bibinfo{person}{Jie Wang}, \bibinfo{person}{Licheng Guo},
  {and} \bibinfo{person}{Jason Cong}.} \bibinfo{year}{2021}\natexlab{}.
\newblock \showarticletitle{AutoSA: A Polyhedral Compiler for High-Performance
  Systolic Arrays on FPGA}. In \bibinfo{booktitle}{\emph{{FPGA}}}.
\newblock


\bibitem[\protect\citeauthoryear{Zuo, Li, Chen, Pouchet, Zhong, and Cong}{Zuo
  et~al\mbox{.}}{2013a}]%
        {zuo2013codegen}
\bibfield{author}{\bibinfo{person}{Wei Zuo}, \bibinfo{person}{Peng Li},
  \bibinfo{person}{Deming Chen}, \bibinfo{person}{Louis-No{\"e}l Pouchet},
  \bibinfo{person}{Shunan Zhong}, {and} \bibinfo{person}{Jason Cong}.}
  \bibinfo{year}{2013}\natexlab{a}.
\newblock \showarticletitle{Improving polyhedral code generation for high-level
  synthesis}. In \bibinfo{booktitle}{\emph{2013 International Conference on
  Hardware/Software Codesign and System Synthesis (CODES+ ISSS)}}. IEEE,
  \bibinfo{pages}{1--10}.
\newblock


\bibitem[\protect\citeauthoryear{Zuo, Liang, Li, Rupnow, Chen, and Cong}{Zuo
  et~al\mbox{.}}{2013b}]%
        {zuo2013improving}
\bibfield{author}{\bibinfo{person}{Wei Zuo}, \bibinfo{person}{Yun Liang},
  \bibinfo{person}{Peng Li}, \bibinfo{person}{Kyle Rupnow},
  \bibinfo{person}{Deming Chen}, {and} \bibinfo{person}{Jason Cong}.}
  \bibinfo{year}{2013}\natexlab{b}.
\newblock \showarticletitle{Improving high level synthesis optimization
  opportunity through polyhedral transformations}. In
  \bibinfo{booktitle}{\emph{{FPGA}}}. \bibinfo{pages}{9--18}.
\newblock


\end{thebibliography}

\end{document}